# Multi-Channel Transfer Learning of Chest X-ray Images for Screening of COVID-19

Sampa Misra[1], Seungwan Jeon[2], Seiyon Lee[1], Ravi Managuli[3], and Chulhong Kim[1,2*]

*Abstract*— The 2019 novel coronavirus (COVID-19) has spread rapidly all over the world and it is affecting the whole society. The current gold standard test for screening COVID-19 patients is the polymerase chain reaction test. However, the COVID-19 test kits are not widely available and time-consuming. Thus, as an alternative, chest X-rays are being considered for quick screening. Since the presentation of COVID-19 in chest X-rays is varied in features and specialization in reading COVID-19 chest X-rays are required thus limiting its use for diagnosis. To address this challenge of reading chest X-rays by radiologists quickly, we present a multi-channel transfer learning model based on ResNet architecture to facilitate the diagnosis of COVID-19 chest X-ray. Three ResNet-based models (Models *a, b*, and *c)* were retrained using *Dataset_A* (1579 normal and 4429 diseased), *Dataset_B* (4245 pneumonia and 1763 non-pneumonia), and *Dataset_C* (184 COVID-19 and 5824 Non-COVID19), respectively, to classify (a) normal or diseased, (b) pneumonia or non-pneumonia, and (c) COVID-19 or non-COVID19. Finally, these three models were ensembled and fine-tuned using *Dataset_D* (1579 normal, 4245 pneumonia, and 184 COVID-19) to classify normal, pneumonia, and COVID-19 cases. Our results show that the ensemble model is more accurate than the single ResNet model, which is also re-trained using *Dataset_D* as it extracts more relevant semantic features for each class. Our approach provides a precision of 94 % and a recall of 100%. Thus, our method could potentially help clinicians in screening patients for COVID-19, thus facilitating immediate triaging and treatment for better outcomes.

This work was supported by the National Research Foundation grant (NRF-2019R1A2C2006269) funded by Ministry of Science and ICT (MSIT), the Republic of Korea, Institute of Information & communications Technology Planning & Evaluation (IITP) grant (No. 2019-0-01906, Artificial Intelligence Graduate School Program) funded by MSIT, the Republic of Korea, Tech Incubator Program for Startup (TIPS) program (S2640139) funded by Ministry of Small and Medium-sized Enterprises and Startups (SMEs) and Technology and Information Promotion Agency (TIPA) for SMEs.

S. Misra, S. Lee, and C. Kim are with Opticho, Pohang 37673, South Korea (e-mail: sampamisra.opticho@gmail.com; ro4797.opticho@gmail.com; chulhong@postech.edu)

S. Jeon, and C. Kim are with the Department of Electrical Engineering,Creative IT Engineering, Mechanical Engineering, and Graduate School of Artificial Intelligence, Pohang University of Science and Technology, Pohang 37673, South Korea (e-mail: jsw777@postech.ac.kr; chulhong@postech.edu).

R. Managuli is with the Department of Bioengineering, University of Washington, Seattle 98195, USA (e-mail: ravim@ u.washington.edu).

## I. Introduction

Coronavirus was detected as a human pathogen in the mid-1960s. It infects humans and a wide range of animals (including birds and mammals). Since 2002, two coronaviruses infecting animals have evolved and caused outbreaks in humans: SARS-CoV (Severe Acute Respiratory Syndrome) identified in Southern China in 2003 and MERS-CoV (Middle East Respiratory Syndrome) identified in Saudi Arabia in 2012 [1, 2]. The new coronavirus disease of 2019 (COVID-19) was first identified in Wuhan, China in December 2019. Since, then the COVID-19 has spread globally, resulting in the ongoing pandemic with the recorded rate that is never seen before for any infectious disease. While the majority of cases result in mild symptoms, many progress to viral pneumonia and multi-organ failure. As of May 6, 2020, more than 3.7 million people across 200 countries have been infected, 0.25 million people have died, and 1.2 million people have been recovered [3]. These numbers change almost every minute. Since the virus is very infectious and spreads through contact and proximity, the whole world is in quarantine significantly affecting the socio-economic welfare of the world population. The real-time reverse transcription-polymerase chain reaction (rRT-PCR) test is the gold standard method of screening the COVID-19 patients [4]. However, this method is slow, not accurate, requires repeated tests, and the test kits are not widely available limiting the number of tests performed. Thus, as an alternative to PCR, widely available chest X-rays are being considered for quick diagnosis or at least stratification of the patient [5]. However, since the COVID-19 chest X-rays are varied in features and presentation, specialization in reading COVID-19 chest X-rays is required thus limiting its use and wider deployment [6]. Hence, it could be of great importance to develop a computer-aided diagnostic (CAD) systems to detect the COVID-19 cases so that it can not only be used to facilitate immediate triaging and treatment of the COVID-19 patients but also to isolate them to prevent the wide and quick spread of the disease.

Among recent developments in deep learning (DL), the convolutional neural network (CNN) has broad applications in computer vision and has opened opportunities for creating a new generation of CAD-like tools for medical imaging [7]. Wang and Wong [8] introduced a deep convolutional neural network based on the ResNet model, called COVID-Net for the detection of COVID-19 cases from open-source chest radiography images. Farooq and Hafeez [9] also used the existing ResNet-50 architecture for the same datasets. Xuanyang et al. [10] developed a data mining technique that is used to distinguish SARS and typical pneumonia based on X-ray images. Several studies have also shown that the presence of COVID-19 can be detected from CT-scans [11-13]. However, CT datasets are limited in number and are not readily available for training using CNN.

Training CNN from scratch is very challenging because it requires a large number of labeled images for better performance, which are difficult to get in this pandemic. The available number of images for COVID-19 is limited as discussed in the above-mentioned literature, limiting the CNN training. Thus, to overcome the limited dataset issue, transfer learning (TL) has been adopted [14]. While using TL, the model is first trained with a large number of annotated natural images from computer vision dataset (ImageNet), and then fine-tuned (optimization) using the X-ray images. Narin et al. [15] developed a three CNN based model using existing TL architecture and reported the highest classification accuracy for the ResNet model. They used chest X-ray images of only 50 COVID-19 patients and 50 normal patients. In this manuscript, we have designed a multi-channel ensemble TL method based on ResNet-18 by combining three different models which are fine-tuned in 3 datasets in such a way that the model can extract more relevant features for each class and hence can identify COVID-19 features more accurately from the X-ray images. Previously, an ensemble model for X-ray images was employed by Chouhan et al. [16] to classify pneumonia and normal images. They combined five different pre-trained models and reported state-of-the-art performance using the ensemble method. Akhani and Sundaram [17] also developed an ensemble method using X-ray images by combining AlexNet and GoogLeNet neural networks to classify the images as TB or healthy.

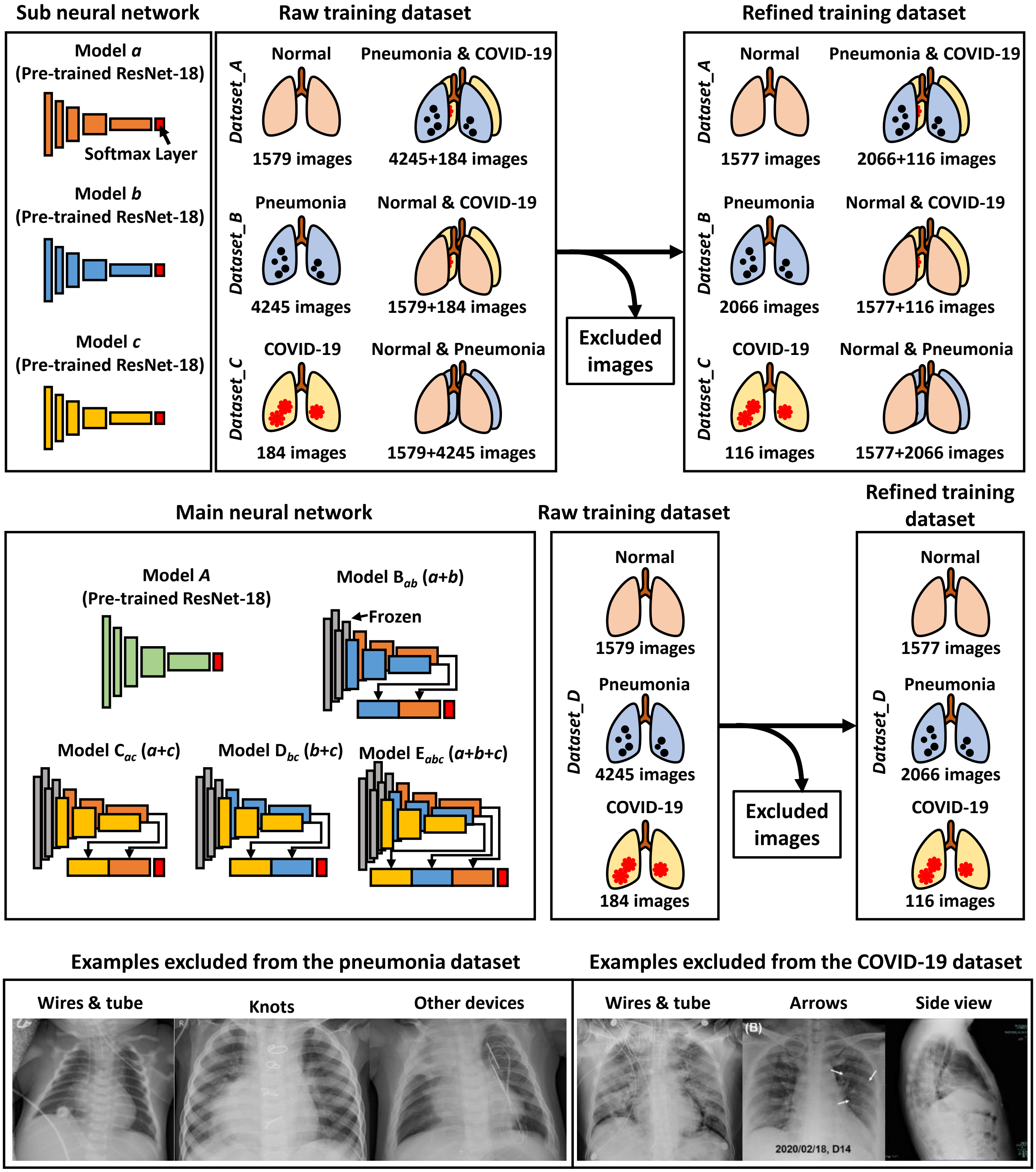

**Fig. 1.** Neural networks and dataset preparation. (a) Sub neural networks (Models *a*, *b*, and *c*) and the training dataset configuration to train the networks. These networks are trained to classify normal or diseased, pneumonia or non-pneumonia, and COVID-19 or non-COVID19 using *Dataset_A*, *Dataset_B*, and *Dataset_C*, respectively. (b) Main neural networks (Models *A*, $B_{ab}$, $C_{ac}$, $D_{bc}$, and $E_{abc}$) and *Dataset_D*. The main neural networks are trained to classify the three cases: normal, pneumonia, and COVID-19. (c) Represented images excluded from the pneumonia (left) and COVID-19 (right) dataset.

## II. Materials and Methods

In this section, first, we will introduce the dataset and deep learning models. Then, we will describe the details of the proposed methodologies.

### A. Dataset & Data Augmentation

We used the chest X-ray images which were adopted from three publicly available X-ray datasets: RSNA Pneumonia Detection Challenge dataset [18], COVID-19 image data collection [19], and COVID-19 X-rays [20]. These datasets are open source and fully accessible to the research community. The first dataset is from Kaggle [18], which consists of 1583 normal and 4273 pneumonia X-ray images. The second (62 images) and third datasets (204 images) are COVID-19 X-ray images from Kaggle [19], and GitHub [20], respectively. However, there were many duplicate images in these sources, and thus after removing the duplicate images the final number of the normal, pneumonia and COVID-19 X-ray images became 1579, 4245, and 184, respectively. Here, we have generated 4 datasets namely *Dataset_A* (normal and disease), *Dataset_B* (pneumonia and non-pneumonia), *Dataset_C* (COVID-19, and non-COVID19), and *Dataset_D* (normal, pneumonia and COVID-19) using these images. The detailed description of the datasets used in this study is presented in Fig. 1. Moreover, we found distinct indicators like markers, wires, and arrows in many images. Thus, we removed the images with distinct indicators and prepared the refined training datasets (Figs. 1a and b). Figure 1c shows the example of deleted images that may overfit the training process.

Data augmentation is commonly used in the medical domain to increase the size of the limited data set [21]. This method generates additional labeled images without changing the semantic meaning of the images. In this manuscript, we used various data augmentation methods like random cropping, rotation, and horizontal flip. In the implementation, the CPU generated the augmented images while the previous batch of images was trained in the GPU. Hence, these data augmentation techniques did not affect the time complexity. We also used oversampling to deal with imbalanced data.

### B. Deep Learning Model

The DL method automatically extracts features from the raw data and then performs the classification of the images. The main advantage of this method is that feature extraction and classification are performed in the same network. The CNN models are the state-of-the-art DL technique, comprising of many stacked convolutional layers to perform automatic feature extraction from image data [22]. It has been used in many radiology applications and is capable of achieving high performance in classifying diseases based on images [7]. The layers used to build CNN architecture are input layer (an image is given as input to produce output), convolutional layers (convolve input image with filters to produces a feature map), rectifier linear unit (ReLU) activation layer (activates the neurons above a threshold value), pooling layers (reduce the size of an image by keeping the high-level features), and fully connected (FC) layers (produce the result). The accuracy of CNN depends on the design of the layers and training data [23]. The CNN generally requires large medical datasets with labels for training, which is difficult to create due to the time and labor cost. Recent studies have shown that transfer learning can be used to overcome this limited dataset size problem.

### C. Transfer Learning

In Transfer Learning (TL), CNN is first trained to learn features in a broad domain e.g. ImageNet. The trained features and network parameters are then transferred to the more specific field. In the CNN model, low label features like edges, curves, corners are learned in the initial layer and the specific high label features are learned in the final layer. Among the different TL models, we chose ResNet since it is well recognized in medical image classification [24]. We employed ResNet-18 because of its shallow architecture, and it can train the images faster without compromising performance. It consists of one 7×7 convolutional layer, 2 pool layers, 5 residual blocks, and one FC layer. The residual block has two 3×3 convolutional layers followed by a batch normalization layer and a ReLU activation function. We skip these two convolution layers and add the input directly before the final ReLU activation function. Classification accuracy of this network is high since this network uses bottleneck residual block, batch normalization (adjust the input layers), and the identity connection (protect the network from vanishing point gradient problem) [24].

### D. Proposed Methodology

The overall proposed methodology is outlined in Fig. 1. We first fine-tuned the CNN architecture that was originally pre-trained (initialized) on the natural image data. Here, 3-ResNet models were fine-tuned based on three different classes (normal, pneumonia, and COVID-19). Model *a* was fine-tuned based on the *Dataset_A* to classify normal from the diseased (i.e., pneumonia and COVID-19) cases. The *Dataset_B* was fed into the Model *b* to identify whether the image is related to pneumonia or not (i.e., normal and COVID-19). Model *c* was fine-tuned based on the *Dataset_C* to classify the COVID-19 cases from non-COVID19 (i.e., normal and pneumonia) cases. These 3 models were pre-trained in parallel to learn respective features to classify normal, pneumonia, and COVID-19 images.

A single ResNet model (Model *A*) that was fine-tuned based on *Dataset_D* (super dataset) that can classify the COVID-19 images from normal and pneumonia images. However, we combined Models *a* & *b*, *a* & *c*, and *b* & *c* to form new Models $B_{ab}$, $C_{ac}$, and $D_{bc}$, respectively (Fig. 1) as ensemble learning allows better prediction compared to a single model. In the ensemble models, every model contributes to the final output, and the weaknesses of the models are offset by the contribution of the other models. Here, the network parameters transferred to the ensemble network and retrained using *Dataset_D* to extract more appropriate features to classify the images as normal, COVID-19, and pneumonia. Finally, we combined these three models *a* & *b* & *c* to construct our proposed Model $E_{abc}$ which extracts various semantic features from the three models. Model $E_{abc}$ can extract more relevant features that can distinguish normal, pneumonia, and COVID-19 more accurately as it is a combination of three specialized sub-models.

The detailed steps for the proposed methodology are as follows:

1. Build Model *a* by fine-tuning the pre-trained ResNet model using the *Dataset_A*, which can classify the normal and

**(a)** **Raw Dataset**

| | Normal | | | Pneumonia | | | COVID-19 | | | Accuracy | MCC |
|---|---|---|---|---|---|---|---|---|---|---|---|
| | Precision | Recall | F1-Score | Precision | Recall | F1-Score | Precision | Recall | F1-Score | | |
| Model *A* (Single ResNet) | 0.776±.11 | 0.944±.07 | 0.844±.05 | 0.984±.02 | 0.884±.07 | 0.940±.03 | 0.838±.07 | 1.000 | 0.910±.04 | 0.903±.04 | 0.801±.06 |
| Model $B_{ab}$ (*a+b*) | 0.830±.04 | 0.964±.02 | 0.892±.01 | 0.988±.01 | 0.920±.02 | 0.950±.01 | 0.896±.04 | 1.000 | 0.944±.02 | 0.933±.01 | 0.858±.01 |
| Model $C_{ac}$ (*a+c*) | 0.812±.01 | 0.980±.01 | 0.882±.01 | 0.996±.01 | 0.902±.01 | 0.948±.01 | 0.894±.10 | 1.000 | 0.940±.06 | 0.924±.01 | 0.850±.03 |
| Model $D_{bc}$ (*b+c*) | 0.846±.10 | 0.942±.07 | 0.884±.04 | 0.984±.02 | 0.918±.06 | 0.950±.03 | 0.893±.12 | 1.000 | 0.928±.06 | 0.940±.07 | 0.870±.01 |
| Model $E_{abc}$ (*a+b+c*) | 0.874±.03 | 0.976±.01 | 0.922±.02 | 0.994±.004 | 0.944±.01 | 0.968±.01 | 0.940±.08 | 1.000 | 0.966± .04 | 0.955±.01 | 0.896±.02 |

**Refined Dataset**

| | Normal | | | Pneumonia | | | COVID-19 | | | Accuracy | MCC |
|---|---|---|---|---|---|---|---|---|---|---|---|
| | Precision | Recall | F1-Score | Precision | Recall | F1-Score | Precision | Recall | F1-Score | | |
| Model *A* (Single ResNet) | 0.874±.06 | 0.950±.04 | 0.910±.01 | 0.970±.03 | 0.878±.06 | 0.916±.02 | 0.674±.05 | 1.000 | 0.804±.03 | 0.909±.01 | 0.830±.02 |
| Model $B_{ab}$ (*a+b*) | 0.876±.02 | 0.982±.01 | 0.924±.01 | 0.982±.01 | 0.882±.02 | 0.930±.01 | 0.834±.08 | 1.000 | 0.910±.04 | 0.927±.01 | 0.862±.01 |
| Model $C_{ac}$ (*a+c*) | 0.884±.03 | 0.980±.01 | 0.928±.01 | 0.984±.01 | 0.880±.03 | 0.930±.01 | 0.734±.06 | 1.000 | 0.844±.04 | 0.924±.01 | 0.862±.02 |
| Model $D_{bc}$ (*b+c*) | 0.896±.01 | 0.970±.01 | 0.932±.004 | 0.978±.01 | 0.896±.01 | 0.936±.005 | 0.694±.06 | 1.000 | 0.820±.04 | 0.930±.005 | 0.868±.01 |
| Model $E_{abc}$ (*a+b+c*) | 0.902±.03 | 0.964±.03 | 0.934±.01 | 0.972±.02 | 0.912±.03 | 0.940±.01 | 0.896±.07 | 1.000 | 0.944±.04 | 0.939±.01 | 0.884±.02 |

**(b)**

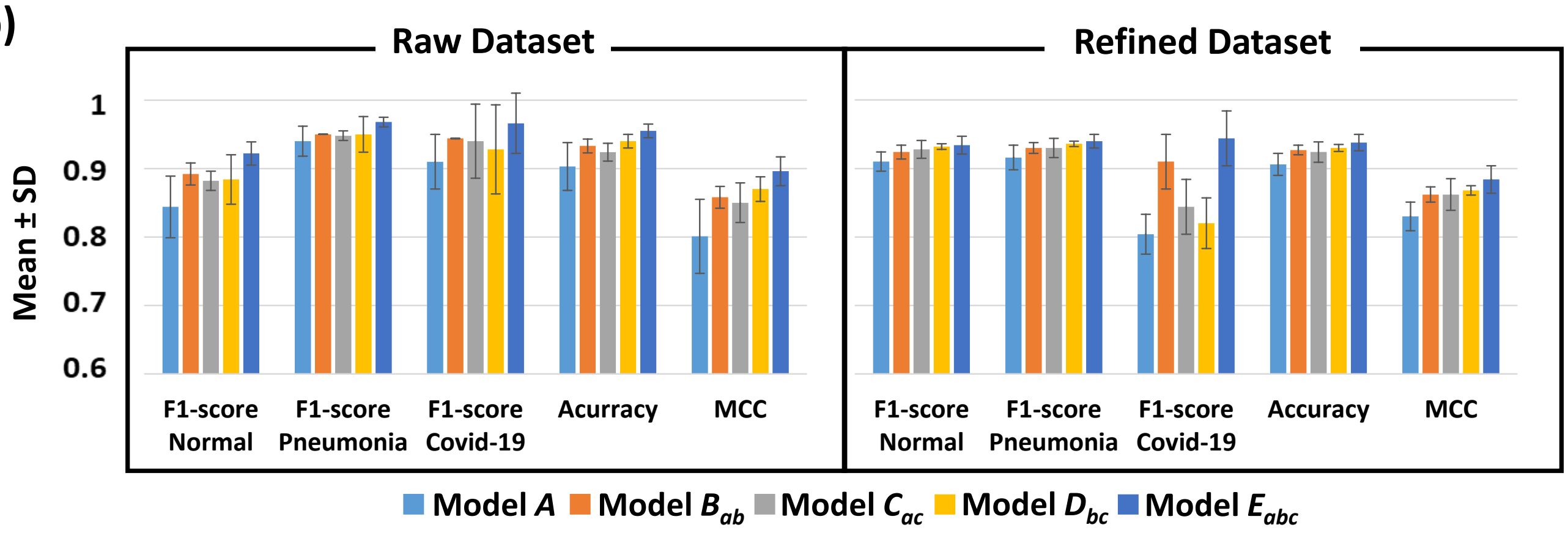

**Fig. 2.** Cross-validated classification performance for each model using both raw and refined datasets. (a) Comparison of the precision, recall, F1-score, accuracy, and Matthews correlation coefficient (MCC). (b) Corresponding graphs that represent each F1-score, accuracy, and MCC of the models trained with the raw and refined datasets.

diseased images.

2. Construct Model *b* to classify pneumonia and non-pneumonia images based on the pre-trained ResNet model by fine-tuning the *Dataset_B*.
3. Design ResNet-based Model *c* by fine-tuning the *Dataset_C*, which can classify the COVID-19 and non-COVID19 images.
4. Remove the classification layer of all models to expose activations of their penultimate layers.
5. Freeze the weights of Models *a*, *b*, and *c*.
6. Build ensemble models (Models $B_{ab}$, $C_{ac}$, $D_{bc}$, and $E_{abc}$) by combining Models *a* & *b*, *a* & *c*, *b* & *c*, and *a* & *b* & *c*.
7. Add a concatenation layer and a classification layer (softmax) into the architecture of the combined models.
8. Train (fine-tune) again the combined models using the *Dataset_D*, which can classify the normal, pneumonia, and COVID-19 images.

## III. Results

In this section, we will describe in detail our experimental setup and results while testing the performance of our method.

### A. Experimental Setup

We trained the deep learning models on a server equipped with a GeForce RTX 2060 super GPU. For training, we used weighted cross-entropy as a loss function and Adam as the optimization function. Here, the weight value for different classes was assigned based on the number of images present in their respective classes. A maximum of 500 epochs was allowed and early stopping criteria was employed, i.e., training was interrupted if the validation loss did not drop in 100 consecutive epochs and the weights of the best epoch were restored.

### B. Classification Performance

The datasets (*Dataset_A*, *Dataset_B*, *Dataset_C*, and *Dataset_D*) were divided into training and testing sets with a ratio of 9:1 in such a way that the images used for test set never used for the training set. We used a five-fold cross-validation scheme to train and measure the performance. We randomly divided the training data into five parts. Among these five parts, four parts were assigned as the training data, and the remaining part was assigned as the validation dataset in such a way that the images used to build the training set were not used for the validation set. This process was repeated five times to train and measure the performance of our proposed method. Then, the averaged values of five folds are used as the final performance values.

We used both raw and refined test set of *Dataset_D* (normal, pneumonia, and COVID-19) to measure the classification performance, where the raw data is the original data without duplicated images and the refined data is the one where duplicates and images with ECG and markings are removed (Fig. 1). The cross-validated performance metrics (i.e., precision, recall, F1-score, accuracy, and Matthews Correlation Coefficient (MCC)) of each model using the raw and refined dataset (*Dataset_D*) are shown in Fig. 2. In Model *A*, a single ResNet was trained to classify three cases simultaneously, with the accuracy and MCC of 90.3% and 0.801 for raw data and 90.9% and 0.83 for refined datasets, respectively. This model has a recall of 100% for COVID-19 but has a low precision value of 67.4% because of poor performance in discriminating pneumonia from COVID-19. All other models that were trained with the ensemble learning exhibit significantly better classification performance than the Model *A*. The ensemble Models $B_{ab}$, $C_{ac}$, and $D_{bc}$ show higher accuracy and MCC than those of the Model *A* by ~2–4% and ~4–7%, respectively, for both raw and refined datasets. From the results, we observe that the ensemble Models $B_{ab}$, $C_{ac}$, and $D_{bc}$ improve the precision compared to the individual Model *A*. It is worth mentioning that the proposed ensemble Model $E_{abc}$, consisting of the three sub-models, has the highest F1-scores, accuracy, and MCC among all models for both raw and refined datasets.

As the decision/classification using a deep learning model is not binary, the predictive probability value (PPV) for each image is calculated and the images are classified as a respective class based on higher PPV value. In order to statistically validate the proposed ensemble methods (Models $B_{ab}$, $C_{ac}$, $D_{bc}$, and $E_{abc}$) with the single Resnet model (Model *A*), *t*-statistics [25] are evaluated using the distributions of those PPV values, which correspond to the true class labels, employing the following expression:

$$t = \frac{\bar{X}_1 - \bar{X}_2}{\sqrt{\left(\frac{(N_1-1)s_1^2 + (N_2-1)s_2^2}{N_1+N_2-2}\right)\left(\frac{1}{N_1}+\frac{1}{N_2}\right)}} \quad (1)$$

Here, ($\bar{X}_1$, $\bar{X}_2$) are the mean and ($s_1$, $s_2$) are the standard deviation of the PPV values for two methods, and ($N_1 = N_2$) is the number of test images. The mean of PPV values, *t* values, and related *p* values for the models are shown in Table 1. The alternative hypothesis implies that the means of PPV values for the ensemble models (Models $B_{ab}$, $C_{ac}$, $D_{bc}$, and $E_{abc}$) are superior as compared to that of Model *A*. The *p* values for the corresponding *t* values eventually reject the null hypothesis that all the approaches are equivalent i.e., they provide the same results with a significance level of 0.05 in favor of the alternative. It is evident from Table 1 that the means of PPV values for ensemble methods are higher than the single ResNet model. It is worth noting that the 3-channel ensemble method (Model $E_{abc}$) has the highest mean of PPV values among all models.

**Table 1.** Results for means of predictive probability value (PPV) and *t*-test evaluated using the distributions of PPV for single ResNet (Model *A*) and ensemble ResNets (Models $B_{ab}$, $C_{ac}$, $D_{bc}$, and $E_{abc}$).

| Models | PPV | *A* | | $E_{abc}$ | |
|---|---|---|---|---|---|
| | | *t*-value | *p*-value | *t*-value | *p*-value |
| *A* | 0.898±.18 | - | - | 4.442 | 0.00001 |
| $B_{ab}$ | 0.915±.16 | 1.684 | 0.046 | 2.184 | 0.025 |
| $C_{ac}$ | 0.918±.20 | 1.913 | 0.028 | 2.145 | 0.016 |
| $D_{bc}$ | 0.918±.20 | 1.913 | 0.028 | 2.145 | 0.016 |
| $E_{abc}$ | 0.942±.16 | 4.442 | 0.00001 | - | - |

## IV. Discussion AND Conclusion

In this manuscript, we present a COVID-multichannel transfer learning method for the classification of patients as normal, COVID-19, and pneumonia based on chest X-ray images. Here, we trained three ResNet-based sub-models (*a*, *b*, and *c*) to classify normal or disease, pneumonia or non-pneumonia, and COVID-19 or Non-COVID19 classes using three different datasets *Dataset_A*, *Dataset_B*, and *Dataset_C*, respectively (Fig. 1a). Then, we combined the sub-models (two or three of these sub-models) and retrained using the *Dataset_D* to extract more appropriate features to classify the images as normal, COVID-19, and pneumonia (Fig. 1b). We observe that the ensemble models (Models $B_{ab}$, $C_{ac}$, $D_{bc}$, and $E_{abc}$) show better classification performance than the single ResNet model (Model *A*) trained on the same dataset. From this, it can be inferred that the superior performance of the proposed method is due to the ensemble learning strategy. It is well known that the ensemble strategy could reduce bias and/or variance in the prediction by utilizing multiple classifiers. In this study, we trained the sub-models with the uniquely configured datasets so that the sub-models *a-c* were trained specifically to screen the normal, pneumonia, and COVID-19 patients, respectively. Due to these specialized sub-models, the ensemble models could extract features to distinguish normal, pneumonia, and COVID-19 more accurately than the single ResNet-based model. The Model's ($B_{ab}$, $C_{ac}$, and $D_{bc}$) classification performance is also improved for the combination of the sub-models. It is worth mentioning that Model $E_{abc}$, consisting of three sub-models, appears to represent the best performance compared to other models (Fig. 2).

To enhance the reliability of deep learning-based screening methods, sufficient consideration should also be given to the composition of datasets. In our result, the recalls for all models are calculated at 100%, and this excessive bias is likely to have a negative impact on other performance metrics. Perhaps, the cause of this problem could be 1) the small number of training images for COVID-19 class, 2) the data separation process, or 3) image artifacts that are present only in that particular class. In this study, the original raw dataset has 184 Covid-19 images, which accounts for only 11.62% and 3.14% of the number of the normal and pneumonia images, respectively. Such a small number of the Covid-19 datasets may not be enough to train the neural networks. In terms of the data separation, the images from the same patient are divided into the training, validation, and test datasets because it is not mentioned whether the images are from the same patient or not in the public dataset we used. Besides, there are many artifacts in the chest X-ray images that may negatively affect the performance of classification tasks for feature-based deep learning models. We have examined the raw datasets manually and there observed many images with wires, arrow signs, etc. (Fig. 1c). These types of images are only present in one particular class, hence it may consider these artifacts as classification features and classify the test images according to these features which are not true features. Therefore, we have deleted these noisy images which are present in one class for the accurate outcome. We observe that the difference of F1-score between raw and refined datasets for COVID-19 class is significantly higher as compared to other classes as there are many noisy images in the original COVID-19 class.

Additionally, we have tested the performance of the proposed ensembled methods using 41 new COVID-19 images (uploaded in the same GitHub repository https://github.com/ieee8023/covid-chestxray-dataset from April 16, 2020-May 5, 2020), 50 normal images and 50 pneumonia images. From Fig. 3, we observed that the ensemble models (Models $B_{ab}$, $C_{ac}$, $D_{bc}$, and $E_{abc}$) show better classification performance than the single ResNet model (Model *A*).

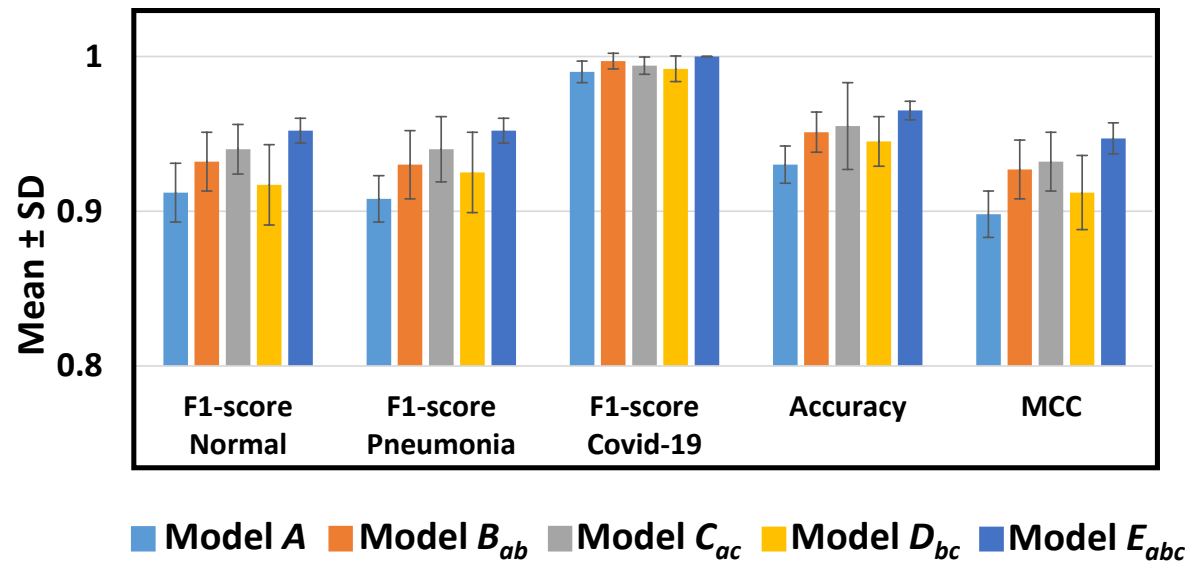

**Fig. 3** Classification performance graph that represent F1-score of normal, pneumonia, and COVID-19, accuracy, and MCC for each model using 41 new COVID-19 images (uploaded in the GitHub https://github.com/ieee8023/covid-chestxray-dataset from April 16, 2020-May 5, 2020)

In conclusion, we propose an ensemble learning strategy to improve the classification performance in deep learning-based COVID-19 screening for chest X-ray images. Moreover, we discuss the potential limitation of the existing open COVID-19 X-ray image dataset by comparing the results before and after refining the dataset. X-ray is a relatively cheap and rapid medical imaging technique, so we can expect its rapid dissemination and high diagnostic throughput, which are important in the urgent pandemic situation. Our deep learning-based method could significantly improve the accuracy of the X-ray COVID-19 screening at a low cost. Accordingly, we believe that this method could enable intensive and efficient support for COVID-19 patients by saving human and temporal medical resources and reduce unnecessary quarantine periods of patients to minimize social cost losses.

## Acknowledgment

S. M., S. L., and C. K. have financial interests in OPTICHO, which supported this work.

## References

[1] M. M. Lai, "SARS virus: the beginning of the unraveling of a new coronavirus," *Journal of biomedical science*, vol. 10, no. 6, pp. 664-675, 2003.

[2] A. Zumla, D. S. Hui, and S. Perlman, "Middle East respiratory syndrome," *The Lancet*, vol. 386, no. 9997, pp. 995-1007, 2015.

[3] W. H. Organization, "Coronavirus disease 2019 (COVID-19): situation report, 72," 2020.

[4] J. F.-W. Chan *et al.*, "Improved molecular diagnosis of COVID-19 by the novel, highly sensitive and specific COVID-19-RdRp/Hel real-time reverse transcription-polymerase chain reaction assay validated in vitro and with clinical specimens," *Journal of Clinical Microbiology*, 2020.

[5] Z. Xu *et al.*, "Pathological findings of COVID-19 associated with acute respiratory distress syndrome," *The Lancet respiratory medicine,* vol. 8, no. 4, pp. 420-422, 2020.

[6] H. X. Bai *et al.*, "Performance of radiologists in differentiating COVID-19 from viral pneumonia on chest CT," *Radiology,* p. 200823, 2020.

[7] G. Litjens *et al.*, "A survey on deep learning in medical image analysis," *Medical image analysis,* vol. 42, pp. 60-88, 2017.

[8] L. Wang and A. Wong, "COVID-Net: A tailored deep convolutional neural network design for detection of COVID-19 cases from chest radiography images," *arXiv preprint arXiv:2003.09871,* 2020.

[9] M. Farooq and A. Hafeez, "COVID-ResNet: A Deep Learning Framework for Screening of COVID19 from Radiographs," *arXiv preprint arXiv:2003.14395,* 2020.

[10] X. Xuanyang, G. Yuchang, W. Shouhong, and L. Xi, "Computer aided detection of SARS based on radiographs data mining," in *2005 IEEE Engineering in Medicine and Biology 27th Annual Conference*, 2006, pp. 7459-7462: IEEE.

[11] F. Shan+ *et al.*, "Lung Infection Quantification of COVID-19 in CT Images with Deep Learning," *arXiv preprint arXiv:2003.04655,* 2020.

[12] O. Gozes *et al.*, "Rapid ai development cycle for the coronavirus (covid-19) pandemic: Initial results for automated detection & patient monitoring using deep learning ct image analysis," *arXiv preprint arXiv:2003.05037,* 2020.

[13] S. Wang *et al.*, "A deep learning algorithm using CT images to screen for Corona Virus Disease (COVID-19)," *medRxiv,* 2020.

[14] S. J. Pan and Q. Yang, "A survey on transfer learning," *IEEE Transactions on knowledge and data engineering,* vol. 22, no. 10, pp. 1345-1359, 2009.

[15] A. Narin, C. Kaya, and Z. Pamuk, "Automatic detection of coronavirus disease (COVID-19) using X-ray images and deep convolutional neural networks," *arXiv preprint arXiv:2003.10849,* 2020.

[16] V. Chouhan *et al.*, "A Novel Transfer Learning Based Approach for Pneumonia Detection in Chest X-ray Images," *Applied Sciences,* vol. 10, no. 2, p. 559, 2020.

[17] P. Lakhani and B. Sundaram, "Deep learning at chest radiography: automated classification of pulmonary tuberculosis by using convolutional neural networks," *Radiology,* vol. 284, no. 2, pp. 574-582, 2017.

[18] R. P. D. Challenge, "Radiological Society of North America," ed, 2018.

[19] Kaggle. Available: https://www.kaggle.com/andrewmvd/convid19-xrays

[20] J. P. Cohen, P. Morrison, and L. Dao, "COVID-19 image data collection," *arXiv preprint arXiv:2003.11597,* 2020.

[21] C. Shorten and T. M. Khoshgoftaar, "A survey on image data augmentation for deep learning," *Journal of Big Data,* vol. 6, no. 1, p. 60, 2019.

[22] R. Yamashita, M. Nishio, R. K. G. Do, and K. Togashi, "Convolutional neural networks: an overview and application in radiology," *Insights into imaging,* vol. 9, no. 4, pp. 611-629, 2018.

[23] W. Rawat and Z. Wang, "Deep convolutional neural networks for image classification: A comprehensive review," *Neural computation,* vol. 29, no. 9, pp. 2352-2449, 2017.

[24] K. He, X. Zhang, S. Ren, and J. Sun, "Deep residual learning for image recognition," in *Proceedings of the IEEE conference on computer vision and pattern recognition*, 2016, pp. 770-778.

[25] T. K. Kim, "T test as a parametric statistic," *Korean journal of anesthesiology,* vol. 68, no. 6, p. 540, 2015.